\documentclass{article}
\usepackage[dblblindworkshop, final]{neurips_2026}
\workshoptitle{The 6th Workshop on Mathematical Reasoning and AI}
\usepackage[utf8]{inputenc}
\usepackage[T1]{fontenc}
\usepackage{hyperref}
\hypersetup{pdftitle={Coding Agents for Coding Theory},pdfauthor={Abraham Yeung}}
\usepackage{url}
\usepackage{booktabs}
\usepackage{graphicx}
\usepackage{float}
\usepackage{amsfonts,amsmath,amssymb}
\usepackage{microtype}

\title{Coding Agents for Coding Theory}

\author{Abraham Yeung\\Stanford University\\\texttt{ayeung16@stanford.edu}}

\setcitestyle{numbers,square}

\begin{document} \maketitle

\begin{abstract}
We spent five weeks using an LLM coding agent on open problems in coding theory: finding large sets of four-letter words, such as DNA barcodes, that stay far apart in edit distance. The agent wrote the verifiers and search code; a human chose the problem and set the verification protocol. Restricting the search to codes with a prescribed symmetry, a classical technique, shrank the problem about fourfold and raised the best known code of length $6$ and minimum edit distance $3$ from $114$ to $120$ words ($E_4(6,3) \ge 120$). The same pipeline improved twelve further lower bounds at lengths $6$ to $9$ and distances $3$ to $6$ (Table~\ref{tab:cells}).

We give the failures equal space. Our own search stopped at $116$ and recorded the last symmetry class as topping out at $112$; a second agent session, running the same search with a better operator, found the $120$. A later verdict that the method did not carry over to length $7$ was wrong for the same reason, and an earlier instance (Section~\ref{sec:audit}) cost three weeks. Each time, an intermediate result was written down, never rechecked, and treated as a fact that ruled out further search. Checking final outputs, as our protocol required, does not catch such errors.
\end{abstract}

\section{Introduction}

LLM coding agents can search large combinatorial spaces. The harder question is whether their output can be trusted. Final constructions can be verified. The intermediate conclusions an agent accumulates along the way, such as which symmetry classes have been searched and which cells are exhausted, usually are not, and those conclusions shape a project's trajectory. This paper describes one five-week project in which agent sessions found genuine improvements to published bounds in coding theory, and also recorded wrong intermediate conclusions that hid one of those improvements for three weeks. We give the failures as much space as the results because, for anyone deciding how to supervise such an agent, the failures are more useful.

\paragraph{The results.} The largest known quaternary code of length $6$ with minimum edit distance $3$, equivalently a single-edit-correcting DNA barcode set, has $120$ codewords. The maintained table of bounds gives $114$--$176$~\cite{houghtentable}, the lower value from a 2011--2012 evolutionary search~\cite{ashlock2012,orth2012}. A recent LLM-guided search reports $113$ for the same problem~\cite{llmguided}. The $120$ comes from formulating the problem as maximum independent set on a $4096$-vertex confusability graph and searching over codes invariant under the Klein four-group acting on the alphabet.

\begin{figure}[H] \centering
\includegraphics[width=\linewidth]{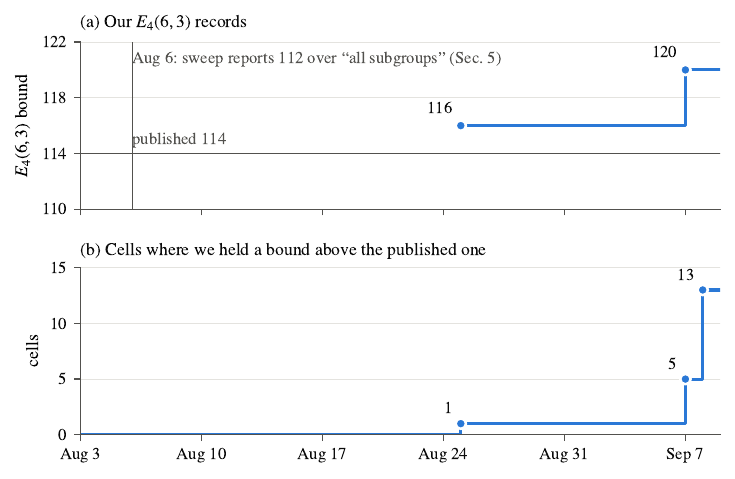}
\caption{When each record was set, by calendar day, from the dated entries in the project notes (\texttt{notes/sota.md}, in the supplement). This is calendar time, not machine time: run records lack consistent timestamps and compute fields, so they cannot be placed on a machine-time axis. (a)~Our $E_4(6,3)$ records against the published $114$; the vertical line marks the sweep of Section~\ref{sec:audit}. (b)~Cells where we held a bound above the published one, at the end of each day. Twelve of the thirteen records date from the last two days, once the ARW operator and the C orbit-graph builder were in place.}
\label{fig:timeline}
\end{figure}

Verification ran in fresh processes: reimplementations sharing no source with the search find a minimum pairwise distance of exactly $3$ and zero colliding edit-balls. The confusability graph was differentially tested against the definition on all $8{,}386{,}560$ vertex pairs, and reproduces the table's exact published entries ($E_4(3,3)=4$, $E_4(4,3)=16$, $n=5$ sweep at the published $44$). Twelve further lower bounds improved (Section~\ref{sec:transfer}, Table~\ref{tab:cells}), at $n=6$ to $9$ and $d$ up to $6$, those at $n \ge 8$ once a C orbit-graph builder made the sizes tractable. Figure~\ref{fig:timeline} (Appendix~\ref{app:timeline}) dates each record.

\section{Setup} \label{sec:setup}

A code $C \subseteq \Sigma^n$ over $|\Sigma|=q$ has minimum edit distance $d$ if every pair of distinct codewords requires at least $d$ insertions, deletions or substitutions to interconvert; $E_q(n,d)$ is the maximum size of such a code. Maximising $E_q(n,d)$ is a maximum independent set (MIS) problem on the \emph{confusability graph}: the vertex set is $\Sigma^n$ ($4^6 = 4096$ vertices at $n=6$), with two words adjacent when their edit distance is less than $d$. At $n=6$, $d=3$ the graph has $392{,}358$ edges, and the strong MIS solver ReduMIS~\cite{redumis} plateaus at $113$.

\paragraph{What the agent did and did not do.} The agent was Claude Opus~5 via Claude Code, an LLM with shell access rather than a bespoke mathematical system. It wrote the verifiers and search code and drove off-the-shelf solvers (HiGHS, CaDiCaL, SciPy/CVXPY, ReduMIS~\cite{redumis}). A human chose the problem, set the protocol and redirected between sessions. The agent did the implementation and interpretation, and chose methods from the protocol's escalation order (Appendix~\ref{app:provenance} lists who proposed what). That protocol, which we call verifier-first, was a project instructions file loaded into every session. It required two independent verifiers with different representations and no shared code, the first written and tested on corrupted objects before any search began. It also required reproducing known published values before searching new cells, and re-verifying every saved bound in a fresh process. In the shipped version, its escalation order puts prescribed-symmetry search right after brute force and moves up a rung when the current one plateaus. The file was rewritten during the project, so we cannot tell whether that advice came before the agent first used the method. The decisive run came from a second agent session with no shared context that audited our repository (Section~\ref{sec:worked}). It used its own implementation of the local search of Andrade, Resende and Werneck (ARW)~\cite{arw}. Our Python operator already had the same (1,2)-swaps and plateau moves, so the difference lay in implementation details we did not isolate. Sessions did not re-read all earlier sessions. This matters for Section~\ref{sec:audit}, because conclusions written in one session became invisible constraints on later ones. The verifiers are independent in representation (edit balls, two Levenshtein tables, Myers bit-vectors) but not in authorship, and the differential test against the definition on all word pairs guards their shared assumptions.

\paragraph{Related work.} The $d=3$ cells were recently attacked by LLM-guided search~\cite{llmguided}. The checks we use are standard testing ideas (known-answer instances, metamorphic and differential testing), normally applied to a system's outputs. What cost this project three weeks was an error in intermediate state and its summaries, which output checks never see.

\section{What worked} \label{sec:worked}

\paragraph{Prescribed symmetry.} The mathematical core is prescribed-automorphism search in the Kramer--Mesner line~\cite{kramermesner}: choose a group, require the code to be invariant under it, and search orbit representatives instead of individual words. At $n=6$ the natural symmetry group is $S_4 \times C_2$ (the $24$ permutations of the four-letter alphabet, times word reversal, giving order $48$). We do not claim this is the full automorphism group of the confusability graph, which we did not compute. The group has $33$ conjugacy classes of subgroups. Twenty-two are product-form (a permutation subgroup $H$, optionally paired with reversal). The other eleven are \emph{diagonal}: they contain elements that pair a permutation with reversal, and cannot be written as $H$ or $H \times C_2$. Three of the order-$4$ classes act freely on $\Sigma^6$. In each, the search collapses from $4096$ words to $1021$ usable orbits (those whose elements are pairwise at edit distance $\ge 3$), and every orbit has exactly $4$ elements, so invariant code sizes are multiples of $4$ and nothing lies between $116$ and $120$.

\paragraph{What our search missed.} Our first sweep covered $30$ of the $33$ classes and gave each four seeds; the other three, missed by our enumeration, were searched later and reach $104$. Two of the three free-acting classes returned $116$; the third returned $112$ on all four seeds. We recorded $112$ as its maximum. Both parts of that recording were wrong.
Re-running the same class with our operator at ten ten-minute seeds returns $116$ on five
of them. Then the second agent session ran its ARW implementation on the same orbit graph. It reached $116$ on
every seed within a minute, and $120$ on two of six seeds. We then reimplemented
that operator in C (\texttt{search/mis\_ils.c}, about $10^3$ times more
iterations per second than our Python). It reproduces the $120$ and finds
$120$s in both $C_4$ classes, so all three free classes contain one. Whether a
seed reaches $120$ is decided early: fifteen ten-minute seeds on the free
classes all returned $116$, while thirty one-minute seeds on the $V_4$ class returned $120$ four
times, so more seeds beat longer ones. Those rates belong to the stronger operator. The original four-seed sweep used
our own operator at $128$ seconds per seed, and that operator never reaches
$120$ at any budget we tried. The sweep was underpowered even for $116$, and its ``tops out at $112$'' was a small sample recorded as settled fact.
Nothing reached $124$. A numerical, uncertified Lov\'asz theta value on the orbit graph, about $45$
orbits ($180$ words), is too loose to say whether a $124$ exists.

The instance reduction is real. Our operator gets $111$ on
the raw graph at $180$ core-minutes and $116$ on the orbit graph at $8.5$, and
the stronger operator gets $120$ on the orbit graph but only $114$ on the raw one. ReduMIS on the same orbit graph returns $116$ on each of three single-threaded ten-minute seeds, against $112$--$113$ on the raw graph, where its seeds stopped on their own after $4.6$--$6.1$ minutes; at the same $30$ core-minutes, the C operator returns $120$ on four of thirty one-minute seeds. So the reduction lifts all three operators to $116$, and the step to $120$ needs a stronger operator on the reduced instance, since neither alone reaches it at $n=6$. The ranking is not
stable in $n$. At $n=8$ ReduMIS reaches $1124$--$1132$ on the $V_4$ orbit graph
against the C operator's $1148$, and at $n=9$ it reaches $3760$ against $3724$
(Table~\ref{tab:cells}).

\paragraph{Calibration cells.} Verification shows that an artifact is valid. It says nothing about whether a search was strong enough for a negative result to mean anything, and most of a project's output is negative results. We propose including in every batch one cell with an independent reference; ours is the unrestricted problem, whose exact value is open, with ReduMIS's $113$ as reference. Table~\ref{tab:calib} shows why: a sweep over $15$ previously unexamined cells returned a plausible spread of values up to $100$, while the same solver scored $99$ on the unrestricted problem, where $113$ is attainable. The batch was measuring the solver rather than the cells, so we discarded it. The check is not sufficient, since the sweep that under-searched the winning class passed its calibration cell. It detects a broken solver but not an under-budgeted one.

\begin{table}[!htbp] \centering \footnotesize \caption{Calibration across three runs, and (right) the compute-response of the repaired operator, nearly flat over $6$--$60$ core-min; the $150$ row is our earlier raw-graph local search, for reference only.} \label{tab:calib} \begin{tabular}{llcccc@{\hskip 1.5em}c} \toprule & \multicolumn{4}{c}{the three runs} & \multicolumn{2}{c}{operator vs.\ budget} \\ \cmidrule(r){2-5}\cmidrule(l){6-7} & run & core-min & best & calib. & budget & calib. \\ \midrule 1 & original operator & \phantom{0}6 & 100 & \phantom{0}99 & \phantom{00}6 (broken op.) & \phantom{0}99 \\ 2 & repaired operator & 60 & 112 & 110 & \phantom{00}6 & 109 \\ 3 & later sweep, same op. & 20 & \textbf{116} & 109 & \phantom{0}15 & 109 \\ &                       &    &              &     & \phantom{0}20 & 109 \\ &                       &    &              &     & \phantom{0}60 & 110 \\ \cmidrule(l){6-7} &                       &    &              &     & 150 \emph{(raw-graph ILS)} & 111 \\ \bottomrule \end{tabular} \end{table}

The check needs a decision rule, and an absolute threshold is the wrong one. Run~3, which produced the $116$, scored $109$, \emph{below} the $110$ that licensed trusting the operator in run~2, so a fixed cutoff would have discarded the run that found the $116$. Normalising by compute avoids this. The right-hand columns plot the repaired operator against budget: $109$ at six core-minutes and $110$ at sixty, a $10\times$ increase buying one codeword. The curve has four points; two share their configuration with runs 2 and 3, so they are not independent of what is being judged. The curve is close to flat, and against it run~1's $99$ at six core-minutes is ten low. So we judge a run by its distance from the compute-response curve instead of from a fixed number. The curve costs four short runs on a known cell.

A cheaper variant needs no known answer. Conjugate subgroups give isomorphic problems, so cells within one conjugacy class must report equal optima, and any spread flags the solver. Run~1 returned $88, 87, 86, 84, 88, 83$ on six isomorphic problems inside a single class.

\section{Where it did not work}

\paragraph{$n=7$: first a negative result, then not.} \label{sec:transfer} Applied unchanged to $E_4(7,3)$, whose record is $356$, the best over $30$ of the $33$ invariance classes was $355$. (Three classes our first enumeration missed, which need three generators, were searched afterwards and reach $340$.) Our notes recorded this as ``the method did not transfer,'' then qualified it as a statement about our search rather than about the cell, since the same operator had already under-searched the $n=6$ winning class.

The qualification was correct. The C operator, on the same orbit graphs, returned $358$ on both of its ten-minute seeds in the order-$6$ class that had given $355$, and $359$ in the reversal-only class ($6844$ orbits). That code consists of $164$ reversal pairs and $31$ palindromes. Each passed all four verifiers and a from-scratch distance check. Three further fifteen-minute seeds of the same operator on the order-$6$ class all gave $355$, the stalling of long seeds seen at $n=6$. Twenty-six further two-minute seeds spread $355$--$364$, and the $364$ is the value in Table~\ref{tab:cells}.
\paragraph{Other distances.} Nothing in the method is specific to $d=3$: symbol permutations and reversal preserve edit distance, so the same $33$ classes act on every conflict graph ``$\mathrm{lev}\le d-1$.'' We built a general-$d$ version (a C graph builder and two independent verifiers with a $3463$-check accept/reject suite) and first checked it on the table's exact cells, which it must reproduce and must not exceed: $E_4(5,3)=44$, $E_4(5,4)=11$, $E_4(6,5)=8$ and $E_4(6,6)=4$ all came back exactly. $E_4(6,4)$, tabled $28$--$32$, then gave $30$ in three classes, and a MaxSAT solve of one of those orbit graphs reports $30$ as the optimum within that class; $E_4(7,4)$, tabled $65$--$128$, gave $72$ in two order-$8$ classes. Each took under an hour including graph construction. The same pipeline then ran at $n=7$ to $9$ and at $d$ up to $6$. The winning class differs by cell (Table~\ref{tab:cells}), and there was no shortcut to trying them all. A third operator, NuMVC, trailed both others everywhere.
\begin{table}[!htbp] \centering \footnotesize
\caption{Every lower bound this project improved, with a symmetry class containing the record
(Appendix~\ref{app:codes} gives each code's full stabiliser), the usable-orbit
count of that class, and how many seeds at the stated budget
reached the value (C operator unless marked ReduMIS). The four bounds above the rule were improved first; the rest came from the same pipeline extended further,
with $n \ge 8$ built by a C orbit-graph builder validated edge-for-edge
against the Python one. Four values are class optima as reported by exact solvers ($(6,4)$ by MaxSAT, the
others by maximum clique of the compatibility graph); none has a checked certificate.}
\label{tab:cells}
\begin{tabular}{lcclrl} \toprule
cell & published & new lower bound & class & orbits & seeds \\ \midrule
$E_4(6,3)$ & 114--176 & \textbf{120} & $V_4$ (free) & 1021 & 8/96 at 1 min \\
$E_4(7,3)$ & 356--614 & \textbf{364} & $C_3 \times C_2$ & 2218 & 2/26 at 2 min \\
$E_4(6,4)$ & 28--32 & \textbf{30} & diagonal, order 4, optimum & 518 & 3/3 at 26 s \\
$E_4(7,4)$ & 65--128 & \textbf{72} & diagonal, order 8 & 1277 & 11/11 at 1--2 min \\ \midrule
$E_4(8,3)$ & 1132--2340 & \textbf{1148} & $V_4$ (free) & 16381 & 1/20 at 10 min \\
$E_4(9,3)$ & 3451--9360 & \textbf{3760} & diagonal $C_4$ (free) & 65531 & ReduMIS, 2/7 at 15--30 min \\
$E_4(8,4)$ & 176--512 & \textbf{200} & diagonal, order 8 & 6500 & 2/24 at 10 min \\
$E_4(9,4)$ & 495--2048 & \textbf{520} & diagonal, order 8 & 28627 & 4/10 at 15 min \\
$E_4(7,5)$ & 18--24 & \textbf{20} & $V_4$ (free), optimum & 3568 & 13/13 at 0.5--1 min \\
$E_4(8,5)$ & 38--96 & \textbf{44} & $V_4$ (free) & 15649 & 2/10 at 10 min \\
$E_4(9,5)$ & 97--384 & \textbf{100} & diagonal, order 8 & 13143 & 4/4 at 15 min \\
$E_4(8,6)$ & 14--32 & \textbf{16} & $V_4$ (free), optimum & 12055 & 8/8 at 5--10 min \\
$E_4(9,6)$ & 27--128 & \textbf{32} & diagonal, order 8, optimum & 5813 & 2/2 at 10 min \\
\bottomrule \end{tabular} \end{table}

\paragraph{SAT and ILP.} These end the protocol's escalation order. For constructions they returned nothing usable. SAT reported \texttt{UNKNOWN} after $462$\,s and $389$\,s on instances known to be feasible (binary deletion codes at $n=11$, from the same project), and the ILP returned unconverged incumbents on the cells here. For certificates the same tools were indispensable: \texttt{drat-trim}-verified UNSAT proofs, including a proof that the $A_4$-invariant optimum is exactly $112$, and LP duality reproducing $24$ of $24$ published LP relaxation values as exact rational certificates. The ranking of tools depends on direction. Heuristic local search built every construction, and formal methods produced every checked certificate. The four class optima of Table~\ref{tab:cells} have none.

\section{The most expensive error} \label{sec:audit}

Three weeks before our first record, the $116$, a sweep reported ``best invariant code $112$, proven optimal in the $A_4$ class'' over what it described as all subgroups. Four defects were behind that sentence. The parameterisation expressed invariance as ``$H \le S_4$, optionally with reversal,'' reaching only the $22$ product-form classes. One of the $116$-word codes found later lives in a diagonal class it could not express. And a size cap skipped $15$ of the $60$ (subgroup, reversal) cells of that parameterisation, precisely the weakest-symmetry ones.

Inside the cap, the MILP returned unproven incumbents on cells it could not finish, and the summary dropped their status. Those last two defects are the instructive ones. Three conjugate cells, isomorphic problems that must share an optimum, were reported as $100$, $92$ and $8$ for a quantity now known to be $\ge 120$. The results file \emph{did} record \texttt{proven\_optimal\_invariant: false} for all three, and the conjugacy check of Section~\ref{sec:worked} would have flagged the spread. (The $A_4$ optimum really is $112$; the error was ``all subgroups.'') The flag was present in the data and absent from the summary, and the summary is what a human collaborator reads. None of the four raised an error. All produced plausible numbers, so a protocol that checks outputs never questioned them. The protocol had the same gap, since the ground-truth verifier did not cover this problem family until ten days after our first record, the $116$. We keep the description ``verifier-first'' because that is what the protocol specified, not what the $116$ received at the time.

Here the agent wrote conclusions faster than its supervisor re-read them, and those conclusions narrowed the search without anyone noticing. No search that trusted the recorded summary would have revisited the class that held the bound. The fix is cheap. Recheck settled conclusions on a schedule, and record unproven solver output as \texttt{unknown} rather than as a value. The protocol had no such rule until 7 September. Our own first correction, which added the diagonal classes, was itself incomplete (it enumerated $30$ of the $33$) and was caught only by separate review sessions.

\section{Limitations}

This is a single case study: one project, one problem family, one human supervisor, and no comparison across projects. The $120$ came from a second agent session rather than our main search, and most later bounds from our C reimplementation of that operator ($E_4(9,3)$ from ReduMIS). The operator comparison of Section~\ref{sec:worked}, which includes ReduMIS on the orbit graphs, and the NuMVC run are the only controls we have. They show that the bounds need both the reduced instance and a strong operator, and that which operator is strongest depends on the cell. We cannot cleanly isolate how much of the result is attributable to the agent versus a conventional search pipeline using the same mathematical machinery. The key algorithmic components (prescribed symmetry, the ARW operator) predate LLMs, and the agent's contribution was the engineering around a classical method, not a new method.

The run records sum to about $130$ core-hours on one laptop. Session counts, API cost and human redirects were not logged, and intermediate conclusions were not counted as they were written, so ``three times'' has no denominator. What we can say is that the three negatives that pruned the search all turned out wrong when re-run at a larger budget, while others, such as the absence of a $124$ at $n=6$, held. The calibration rule has no fixed tolerance and has only been tried on the runs in Table~\ref{tab:calib}, and the remedies themselves are untested. The failure modes are not specific to language models. Dropped provenance and underpowered sampling happen in any computational project, and our claim is only that here the agent wrote summaries faster than its supervisor re-read the logs. Every result is an improved lower bound, no upper bound changed, and the incumbents are unverified heuristic claims whose codeword lists were never published, so their long standing may partly reflect how few have attacked them. The logs are self-reported by the system under study, so we release them: the supplement (\url{https://github.com/Abraham-y/coding-agents-coding-theory}) ships the codes, all four verifiers, the defective sweep, the protocol document and every run record, and one script re-verifies every claimed bound in minutes.

\section{Conclusion}

The mathematics that produced the bounds is classical prescribed-automorphism search. Which operator and symmetry class worked best changed from one $(n,d)$ to the next, and the free action fixed sizes at multiples of four without causing the escape from the incumbent. The contribution to AI methodology is the failure analysis. Three
times in this project a conclusion was recorded, never rechecked, and turned out
to be hiding the answer: a sweep whose summary dropped the solver's ``unproven''
status, a four-seed result read as a property of a class, and a ``does not
transfer'' verdict that was a property of the operator. A verifier-first
protocol catches errors in final outputs but not in the intermediate state that decides which outputs to pursue, and we expect that gap to matter more as agents run for longer.

\appendix
\section{Who proposed what} \label{app:provenance}
Compiled from Sections~\ref{sec:setup}--\ref{sec:audit} and the project notes. ``Agent'' means the main agent sessions; ``second session'' means the independent session that audited our repository and found the $120$; ``review sessions'' are other independent adversarial-review sessions.

\begin{table}[!htbp] \centering \small
\begin{tabular}{p{0.52\linewidth}p{0.42\linewidth}} \toprule
Decision or component & Source \\ \midrule
Problem: open cells of the quaternary edit-metric table & Human \\
Protocol: verifier-first rules and escalation order & Human \\
Redirects between sessions & Human \\
Prescribed-symmetry search, a classical method~\cite{kramermesner} & Agent, within the protocol's escalation order, which as shipped lists it second; we cannot date that entry against the agent's first use \\
Verifiers, search code, orbit-graph builders, and C implementations of the ARW operator and NuMVC & Agent \\
Recorded conclusions, including the $112$ ceiling and ``did not transfer'' & Agent \\
Its own ARW implementation~\cite{arw} on the $n=6$ orbit graph, which found the $120$ & Second session \\
Catching that our first correction enumerated only $30$ of the $33$ classes & Review sessions \\
DRAT proof that the $A_4$-invariant optimum is $112$ & Review session \\
Solvers: HiGHS, CaDiCaL, drat-trim, RC2 (python-sat), SciPy/CVXPY with SCS, networkx, ReduMIS~\cite{redumis} & Existing tools \\
\bottomrule \end{tabular} \end{table}

\clearpage
\section{Timeline} \label{app:timeline}

\section{The thirteen codes} \label{app:codes}
Each list is the complete code; sha256 is over the sorted words, one per line, each line newline-terminated. Each was re-verified from the definition in a fresh process (pairwise Levenshtein distance $\ge d$). The same lists ship as plain-text files in the supplement, with the verifiers.

\paragraph{$E_4(6,3) \ge 120$, 120 words.} Stabiliser in $S_4\times C_2$: Klein four-group $V_4$ on symbols, no reversal. sha256 {\footnotesize\texttt{adeaa3604074f5f2b42ece623e45df96f9521650b98fb11a7182321fbc5c3576}}.

{\small\ttfamily\sloppy 000323 001231 001302 002011 002103 003000 003222 010022 010110 011223 011311 012132 012300 013033 013201 020312 021021 021113 022030 022202 023131 023303 030013 030220 031032 031101 032233 032321 033112 033330 100200 100332 101001 101133 102122 102310 103023 103211 110213 110320 111232 112111 112333 113012 113100 120010 120123 121102 121331 122003 122221 123230 123322 130002 130130 131203 132020 132212 133121 133313 200020 200212 201121 201313 202130 203203 203331 210011 210103 211112 211330 212002 212231 213210 213323 220233 220321 221000 221222 222101 223013 223120 230122 230310 231023 231211 232200 232332 233001 233133 300003 300221 301012 301100 302232 302301 303113 303320 310030 310202 311131 311303 312220 312312 313021 320132 320300 321033 321201 322022 322110 323223 323311 330111 330333 331230 331322 332031 332102 333010\par}

\paragraph{$E_4(7,3) \ge 364$, 364 words.} Stabiliser in $S_4\times C_2$: $C_3 \times C_2$, the 3-cycle $(1\,2\,3)$ on symbols and reversal. sha256 {\footnotesize\texttt{dadc469672cd5dfe21dbe551ca884c81851a9cc661440f323cf7c65d895e34e6}}.

{\small\ttfamily\sloppy 0000000 0001331 0002112 0003223 0010113 0011100 0011312 0012233 0012321 0013010 0013202 0020221 0021020 0021303 0022123 0022200 0023132 0023311 0030332 0031122 0031213 0032030 0032101 0033231 0033300 0100123 0100330 0101032 0103100 0103212 0110020 0111121 0111203 0112222 0112313 0113230 0120210 0121130 0122111 0122320 0123022 0130112 0130201 0131220 0131302 0132133 0133003 0133321 0200110 0200231 0201200 0201323 0202013 0210223 0210302 0211001 0211132 0212103 0212330 0213211 0220030 0221310 0222232 0222301 0223121 0223333 0230320 0231033 0232210 0233130 0233222 0300220 0300312 0302131 0302300 0303021 0310130 0311210 0311333 0312011 0313320 0320103 0320331 0321322 0322002 0322213 0323110 0323201 0330010 0331111 0331232 0332120 0333102 0333313 1000213 1000302 1001120 1002001 1003131 1010101 1011211 1012132 1012300 1013033 1020003 1020122 1020310 1021133 1022212 1023021 1023230 1031023 1032220 1032313 1033112 1100011 1101113 1101202 1102130 1103333 1110233 1110321 1111022 1111330 1112123 1112210 1113312 1120032 1121101 1122013 1122302 1123120 1123211 1130103 1130222 1131132 1131311 1132031 1133200 1200133 1201021 1203030 1203201 1203322 1210031 1211110 1213102 1213220 1220200 1220323 1221332 1222022 1222131 1223113 1230002 1230111 1231212 1231303 1232100 1232321 1233233 1233310 1300121 1301223 1302102 1302311 1310332 1311103 1312030 1312221 1313001 1313213 1320020 1321112 1321231 1322333 1323222 1323300 1330230 1331000 1332012 1332203 1333123 1333331 2000103 2000321 2001212 2002230 2003002 2011223 2012031 2013121 2013330 2020202 2021011 2022322 2023100 2023213 2030001 2030120 2030233 2031032 2031310 2032211 2033323 2100232 2102331 2103122 2103203 2110310 2111112 2111231 2112000 2113023 2113301 2120113 2121002 2121321 2122201 2123010 2123332 2130030 2131100 2131333 2132223 2132312 2133111 2200022 2201111 2202221 2202303 2203210 2210201 2210333 2211300 2212122 2212213 2213012 2220132 2220311 2221123 2222033 2222110 2223231 2223320 2230013 2231230 2231322 2232202 2233021 2233103 2300211 2301010 2301133 2301302 2302032 2310003 2310222 2311120 2311311 2312101 2312323 2313132 2313200 2320012 2321203 2321330 2322220 2330131 2330300 2331221 2332113 2333033 2333212 3000132 3000201 3001003 3002323 3003310 3010002 3010230 3010311 3011131 3012013 3012120 3013322 3021110 3021232 3022331 3023012 3030303 3031200 3031321 3032022 3033133 3100322 3102020 3102103 3102211 3103013 3110100 3110212 3111011 3111323 3112332 3113221 3120001 3120333 3121213 3121300 3122122 3122230 3123131 3123202 3130023 3132110 3132301 3133330 3200313 3201233 3201301 3203112 3210010 3211222 3212111 3212200 3213123 3213331 3220120 3221031 3221102 3222223 3222312 3223000 3230221 3231020 3231113 3232003 3232132 3233302 3300033 3301320 3302222 3303101 3303332 3310021 3311032 3311201 3312133 3312310 3313303 3320111 3320302 3321023 3322100 3323233 3323321 3330122 3330213 3331130 3331312 3332231 3333011 3333220\par}

\paragraph{$E_4(8,3) \ge 1148$, 1148 words.} Stabiliser in $S_4\times C_2$: Klein four-group $V_4$ on symbols, no reversal. sha256 {\footnotesize\texttt{e98fe9d121189c38208634f4b8978be329f6aeee0f2d2bedd97105bb87c57fb8}}.

{\small\ttfamily\sloppy 00000010 00000221 00002031 00002322 00003101 00003313 00010323 00011003 00011211 00012130 00013122 00020332 00021202 00022213 00023020 00030131 00031320 00032103 00032232 00033012 00100002 00101112 00101321 00102123 00102310 00103200 00103332 00110233 00111302 00112020 00112212 00113013 00120103 00121000 00122333 00123221 00130201 00131230 00132111 00133303 00200120 00201232 00202211 00203023 00210011 00210222 00211121 00211333 00212203 00213312 00220210 00221113 00221322 00222001 00223032 00223300 00230112 00231002 00232033 00232220 00233231 00300113 00301203 00301330 00302100 00303212 00310030 00310101 00312313 00313110 00320311 00322131 00322222 00323003 00323230 00330223 00330300 00331031 00331122 00333133 01000103 01001200 01003211 01010012 01011232 01013100 01020233 01021030 01022003 01022120 01023322 01030001 01031302 01032022 01033033 01033121 01101213 01103003 01110023 01110111 01110200 01111122 01121133 01122232 01122311 01123031 01130212 01130320 01131331 01132221 01133010 01200223 01201301 01202020 01203122 01203330 01211013 01213220 01220132 01221021 01221100 01223212 01230030 01231123 01231310 01232011 01232332 01233203 01300210 01300331 01301022 01301111 01302133 01302302 01303323 01310120 01311103 01311321 01312001 01313332 01320303 01321223 01321312 01322200 01330013 01331000 01333102 01333311 02001002 02002111 02003021 02011300 02012033 02012102 02013131 02020212 02021103 02021321 02022310 02023200 02023333 02030032 02030220 02030313 02031112 02100232 02101120 02102201 02103030 02103113 02111032 02113321 02120010 02121211 02122022 02123302 02130123 02131101 02131222 02132300 02133233 02200200 02200311 02201133 02202222 02202303 02210130 02211223 02212012 02212320 02213000 02220002 02220231 02221330 02222121 02223011 02231020 02232113 02233132 02233210 02233301 02300020 02302213 02303310 02310312 02311202 02312331 02313211 02321001 02322030 02322112 02322323 02330110 02330321 02331333 02332002 02333023 03000312 03001303 03002023 03002132 03003000 03011331 03012113 03012200 03013011 03013223 03020022 03020111 03021210 03022231 03023130 03031013 03031221 03032301 03033332 03100110 03100333 03101231 03110221 03111001 03111313 03112322 03113112 03113330 03120300 03121012 03122203 03123023 03130132 03132031 03132210 03133002 03133120 03200003 03200131 03201320 03203233 03210020 03210213 03210332 03212311 03213103 03220123 03222110 03222302 03223201 03230202 03231111 03232000 03232122 03233313 03300122 03300201 03302220 03303032 03311010 03312233 03320033 03321102 03322011 03323213 03323320 03330103 03331130 03332312 03333222 10000033 10001000 10001132 10001311 10010302 10012112 10020220 10021231 10021303 10022101 10023330 10030022 10032120 10033200 10033323 10100323 10101103 10102011 10110311 10111012 10112300 10120213 10121110 10122030 10122122 10123133 10130121 10131322 10132233 10133031 10133112 10200012 10200230 10201031 10202223 10203110 10210000 10210133 10211220 10211301 10212232 10213022 10213213 10220111 10221102 10222013 10222200 10230203 10230332 10231212 10233311 10300102 10302331 10310210 10311332 10312033 10312221 10313131 10320032 10320201 10321121 10322312 10323100 10323223 10330011 10330130 10331023 10332303 11000213 11001322 11002102 11003131 11003303 11010003 11010230 11011113 11012223 11012311 11013032 11013201 11020321 11021310 11022212 11023000 11030111 11031012 11032330 11033222 11100112 11100300 11101232 11102033 11103021 11111101 11111330 11112010 11112202 11113120 11113233 11120231 11121020 11122103 11123012 11123323 11130313 11131223 11132131 11133302 11201010 11201121 11202001 11203202 11210221 11210312 11211002 11212303 11213011 11220033 11220120 11221211 11221332 11222022 11231200 11232112 11232321 11233020 11233333 11300030 11300222 11301100 11301333 11302203 11303312 11310323 11311031 11312132 11313300 11320113 11321003 11322320 11323122 11323331 11330002 11330233 11331301 11332123 11332211 11333110 12000110 12000202 12001330 12002003 12002221 12003233 12010320 12011001 12011222 12021023 12022031 12022113 12023120 12023301 12030103 12031211 12032132 12033312 12100220 12102100 12102332 12103002 12103311 12110212 12111203 12112111 12113023 12113132 12120102 12120330 12122223 12123210 12130301 12131000 12131133 12132021 12133320 12200101 12203322 12211033 12211310 12212123 12213331 12220021 12221012 12222333 12223203 12230013 12231122 12232231 12232302 12233100 12301131 12301223 12301302 12302012 12303200 12310231 12311020 12311112 12312322 12320000 12321313 12322202 12323033 12323111 12331032 12332310 12333001 12333213 13000123 13002230 13010031 13011323 13012002 13012121 13013310 13020010 13020333 13021032 13022322 13023211 13030300 13031101 13032213 13033133 13100211 13102020 13103013 13103122 13110113 13112130 13113000 13120003 13121123 13121202 13121331 13130012 13130230 13131303 13132222 13132311 13133201 13200313 13201203 13202300 13203220 13211131 13212201 13213302 13220222 13221001 13221230 13222132 13223113 13230110 13233003 13233121 13233232 13300332 13301021 13302111 13303103 13303231 13310022 13311200 13311311 13313212 13313333 13320131 13322023 13322210 13322301 13323002 13330221 13331113 13331320 13332100 13333030 20000303 20001233 20002013 20002220 20003112 20010331 20011032 20011123 20011310 20013202 20020000 20020121 20022022 20022133 20023311 20030102 20030230 20031222 20032312 20033001 20100101 20100212 20100330 20103223 20110220 20111201 20112103 20112332 20113111 20120031 20121132 20122202 20130113 20131033 20132130 20133020 20200132 20201022 20201111 20202030 20203103 20203321 20212002 20212131 20212210 20213330 20220333 20221203 20223220 20230211 20230320 20231313 20233122 20300200 20301120 20302232 20303033 20310122 20311011 20312301 20313000 20313323 20320023 20321212 20321331 20322010 20323302 20331103 20333210 21000120 21000332 21001023 21002301 21010222 21010300 21011131 21012020 21013333 21021011 21022221 21022313 21023102 21030133 21031321 21032031 21032110 21032202 21100233 21101031 21101102 21102211 21103320 21110130 21111000 21112321 21113312 21120002 21121210 21122023 21122300 21130011 21133232 21200013 21201312 21202200 21202333 21203032 21210123 21211110 21213003 21213231 21220201 21220310 21221222 21222130 21223121 21230022 21230331 21231001 21231233 21233113 21300021 21301201 21302122 21303230 21310032 21310213 21311220 21311302 21312310 21322111 21322332 21323013 21330100 21331112 21331330 21332003 21333131 21333223 22000223 22001122 22001210 22002032 22003100 22003331 22010002 22010211 22011013 22012330 22013220 22020033 22021201 22022302 22023010 22030021 22031130 22032000 22032233 22033111 22033303 22100000 22100313 22101012 22101221 22102133 22111311 22112001 22112122 22113213 22113300 22120322 22121030 22122331 22123021 22123112 22130131 22131332 22132212 22132323 22200031 22201202 22202110 22203020 22210010 22210321 22211230 22212313 22213102 22220100 22220213 22221131 22221323 22222003 22222232 22230312 22231300 22232101 22233033 22233221 22300111 22301003 22302321 22303222 22310333 22311121 22312023 22313012 22320132 22320301 22321022 22321110 22322220 22323103 22323330 22330030 22330202 22331231 22332011 22333120 23001030 23002310 23003203 23003322 23010110 23010233 23011021 23012212 23013132 23013301 23020202 23021112 23021300 23022001 23023123 23031002 23033231 23100022 23102121 23103001 23103130 23111133 23111320 23112231 23113222 23120120 23120311 23121101 23122032 23122113 23123200 23123333 23130223 23131110 23132302 23133103 23133321 23200221 23200302 23201100 23202011 23203212 23210200 23211211 23211303 23212223 23213120 23221033 23222321 23223022 23231322 23232230 23233010 23300010 23300133 23301213 23303311 23310003 23311232 23312030 23312102 23313113 23321221 23323031 23332022 23332201 23332333 23333300 30000111 30001021 30002203 30003230 30010013 30010120 30011322 30012231 30013300 30021100 30022323 30030301 30031113 30033132 30033211 30100020 30101211 30101333 30102222 30103131 30110132 30111031 30111223 30113210 30120230 30121022 30123001 30123120 30123313 30130100 30132013 30133202 30133330 30200213 30200331 30201123 30201302 30203201 30210310 30211130 30212321 30213033 30220003 30220221 30221011 30222020 30222332 30223112 30232102 30233000 30233223 30300001 30301032 30302112 30302320 30310203 30311102 30312123 30313222 30313311 30320110 30320322 30321133 30321220 30322002 30330333 30331201 30331310 30332030 30333021 31000310 31001331 31002000 31003012 31003223 31011010 31011221 31011303 31012332 31020122 31021002 31022131 31023021 31030023 31031120 31033313 31100032 31100123 31100201 31101220 31102313 31110322 31111212 31112003 31113102 31113331 31120333 31121013 31121321 31122110 31123203 31131030 31131111 31132200 31133022 31133133 31200100 31201033 31202111 31202232 31203210 31210031 31211311 31212122 31213323 31220012 31222301 31230220 31230303 31231132 31232213 31233101 31302221 31303020 31303113 31303301 31310000 31310133 31311023 31312012 31312230 31313121 31320202 31321231 31321300 31322033 31330312 31331222 31332331 32000022 32000231 32002333 32003320 32011133 32012021 32012110 32013030 32020001 32021332 32022012 32022230 32023213 32030010 32031031 32031200 32032222 32032311 32033002 32033123 32100130 32101001 32101322 32102023 32102210 32103303 32110121 32112233 32112312 32113201 32120113 32122320 32130003 32130211 32131313 32132032 32133110 32200323 32201112 32202002 32203013 32203121 32210302 32211022 32211101 32212200 32222211 32223133 32223222 32223310 32230330 32232120 32300212 32300300 32301311 32302031 32303332 32310011 32311213 32311330 32312303 32313100 32320233 32322101 32323321 32330122 32332133 32333230 33000200 33002211 33002302 33003033 33003110 33010103 33010330 33011111 33011202 33013022 33020223 33021020 33023232 33023303 33030121 33031233 33032003 33032130 33033220 33100102 33101113 33101300 33102331 33103221 33110033 33110301 33111332 33112011 33112220 33113123 33120021 33121130 33122000 33122212 33123111 33123322 33130310 33131122 33132101 33133213 33200030 33201222 33202103 33203132 33210112 33211000 33212333 33213230 33220320 33221121 33221313 33222031 33223100 33230001 33230133 33231023 33231210 33232012 33232221 33233331 33300321 33301101 33301230 33302013 33303202 33310313 33311120 33312131 33313001 33320211 33321203 33322122 33322330 33323010 33330020 33330232 33331011 33331302 33333112 33333323\par}

\paragraph{$E_4(9,3) \ge 3760$, 3760 words.} Stabiliser in $S_4\times C_2$: the diagonal $C_4$ generated by $((0\,2\,1\,3),\mathrm{rev})$; found by ReduMIS. sha256 {\footnotesize\texttt{9c7c0ce7c75fbf30820c6af6863e58b28b867e0320ff2476abe864f29bcffb17}}.

{\small\ttfamily\sloppy 000000200 000000313 000002032 000003121 000011010 000012113 000013201 000013332 000020222 000021102 000022203 000022312 000023020 000031131 000031223 000032211 000033310 000100301 000101003 000102330 000103322 000110202 000111122 000111313 000112233 000120123 000121331 000122001 000122220 000123032 000123100 000123211 000130021 000131212 000133133 000202000 000210023 000210130 000210312 000211001 000213111 000220210 000221133 000222231 000223301 000231230 000232322 000233003 000300132 000301020 000302213 000303002 000303110 000303333 000310000 000310211 000311032 000320033 000321323 000323131 000330331 000332023 001000102 001001323 001002001 001002310 001010333 001011120 001012223 001013013 001021300 001023112 001030321 001032122 001100320 001101231 001102013 001110031 001110210 001111023 001113121 001120101 001123203 001131011 001132020 001132333 001133222 001133300 001200221 001200303 001201113 001201202 001203210 001212230 001213022 001213133 001220232 001220311 001221010 001222213 001232301 001233102 001233323 001301130 001302123 001303301 001310302 001311112 001312201 001313310 001320012 001321003 001322111 001322332 001323030 001330103 001330220 001331121 001333211 002000022 002002131 002003100 002003311 002010111 002011330 002012212 002013233 002020230 002022321 002023302 002030031 002031012 002032220 002033132 002100103 002101132 002102210 002103201 002110300 002112003 002112221 002113012 002113331 002120332 002121021 002121310 002122113 002130211 002131120 002132312 002201033 002201312 002202011 002210222 002212123 002213030 002221002 002222333 002223023 002223200 002230013 002231203 002231331 002233212 002300030 002302102 002302320 002310133 002321122 002321211 002322130 002322223 002323313 002330112 002332010 002332232 002332303 002333101 003000133 003000211 003001112 003001230 003002023 003003222 003003303 003010310 003011103 003012121 003020331 003021313 003022232 003023011 003031322 003032330 003033000 003033231 003101022 003101213 003102120 003103030 003111200 003111332 003120002 003120233 003121111 003122300 003123321 003131301 003132102 003133313 003202203 003203021 003211231 003212010 003212302 003213320 003220030 003221220 003222022 003222101 003230122 003230200 003231023 003231110 003232133 003232221 003300101 003301311 003302012 003310021 003311222 003313230 003320113 003321031 003321100 003323212 003331132 003332001 003333033 003333120 003333302 010000221 010001322 010003133 010010012 010011033 010011220 010020311 010021021 010030030 010030202 010032110 010033013 010033321 010101032 010101210 010102020 010103223 010103311 010110111 010112331 010113212 010121101 010122130 010131303 010132011 010133330 010200011 010200330 010201023 010203231 010211211 010212100 010213313 010220223 010221110 010223002 010223121 010230131 010231332 010232212 010300233 010301102 010302111 010302200 010310031 010310222 010311123 010312132 010320303 010321000 010322322 010331012 010331221 010332313 010333020 011000300 011001011 011001233 011002113 011003023 011011131 011011302 011013100 011022031 011022120 011023232 011023301 011030132 011032002 011101123 011101301 011110122 011110201 011112213 011120003 011121212 011121320 011123011 011123102 011123333 011130310 011132103 011132230 011132322 011200020 011202211 011203112 011210323 011211232 011211310 011213003 011213331 011220302 011221001 011222012 011222133 011230100 011231013 011231321 011233032 011300002 011302033 011302222 011303121 011303230 011303313 011310010 011320231 011321113 011322203 011330111 011331030 011332331 011333202 012000130 012001331 012003012 012010001 012011122 012012231 012012303 012013020 012020113 012021323 012022332 012031201 012100222 012102300 012111312 012113033 012120010 012120131 012122022 012130301 012131023 012132121 012133200 012202202 012203123 012210032 012210213 012211133 012211301 012212112 012221030 012222221 012223303 012230230 012231222 012232033 012233001 012300211 012301022 012303132 012310121 012310330 012311200 012313002 012313223 012321232 012322011 012323100 012330203 012331110 012331333 012333322 013000010 013001100 013002201 013002312 013011311 013012030 013013210 013020101 013021132 013022213 013023320 013030123 013032131 013033333 013100200 013102333 013103113 013110223 013111110 013112101 013113300 013120322 013121203 013122112 013122231 013131122 013133012 013133221 013200313 013201212 013202032 013202110 013203302 013210102 013212021 013213222 013220120 013221333 013223311 013231031 013232000 013232323 013233130 013300331 013303000 013311020 013312013 013312202 013313131 013320023 013322310 013323201 013323332 013330232 013331101 013332220 013333213 020000302 020001311 020002003 020002210 020010110 020010323 020011132 020013000 020022022 020023033 020023212 020032300 020033130 020100000 020100122 020100211 020101133 020102031 020103303 020110033 020110220 020111301 020112012 020121030 020122323 020123021 020131110 020131231 020132132 020132203 020133312 020201200 020202112 020202223 020203020 020210232 020212201 020212333 020220102 020220331 020221011 020221322 020230001 020230123 020232030 020232311 020233213 020300320 020303232 020310103 020311021 020311230 020311312 020321213 020322110 020330210 020331202 020333111 021000120 021002231 021003101 021003333 021011001 021012033 021013322 021020111 021021312 021030330 021033123 021100230 021102110 021103212 021110332 021112222 021112303 021113130 021120213 021121022 021121201 021122311 021130131 021130202 021132001 021200132 021201330 021202302 021203203 021210113 021211012 021211221 021221120 021221303 021222023 021223321 021230033 021231000 021231111 021233310 021300013 021301122 021312011 021313032 021320000 021320333 021322102 021322230 021330021 021331320 021332213 022000112 022000201 022001030 022001222 022002123 022011113 022012102 022013221 022020023 022021100 022022001 022023310 022030010 022030303 022031332 022032111 022102101 022103002 022103233 022110311 022111010 022111121 022111202 022113103 022113320 022120200 022121003 022122212 022122330 022130223 022131300 022133031 022133122 022200121 022201103 022203013 022203300 022210000 022212321 022213332 022220130 022220211 022221233 022231021 022232012 022233133 022233220 022301011 022302000 022302313 022303321 022310322 022311033 022312210 022313112 022320031 022321020 022323222 022330120 022331131 022333003 022333330 023000223 023001002 023001121 023010003 023010230 023013120 023013313 023020012 023021330 023022221 023022303 023031133 023031211 023032013 023032202 023033112 023033301 023100021 023101220 023103332 023111031 023112123 023113211 023121302 023122033 023123101 023130030 023132321 023133023 023200322 023201010 023202131 023203230 023210301 023211130 023211323 023212220 023213033 023220203 023220310 023222113 023223000 023223122 023230212 023231102 023233011 023300202 023301233 023303103 023310213 023311201 023312100 023313010 023320121 023321112 023323130 023323323 023330300 023331313 023332022 023332231 030002331 030003022 030003230 030010122 030010213 030010300 030012221 030013103 030013311 030020332 030021123 030021211 030022100 030030023 030030111 030031001 030031232 030032012 030100232 030101111 030101302 030111223 030111330 030112102 030120031 030120112 030121200 030123013 030131321 030133002 030200113 030200201 030202130 030211131 030211303 030213220 030221032 030222003 030222222 030230233 030231010 030300012 030303211 030311110 030312022 030312203 030313323 030320310 030322121 030322333 030323001 030330130 030331103 030332302 030333222 031000032 031001212 031003221 031011230 031011313 031012010 031020133 031021002 031022330 031030203 031033031 031100100 031100312 031102321 031103033 031110013 031111133 031111322 031112032 031113020 031120222 031130123 031131210 031133101 031133332 031201311 031202103 031203000 031210030 031211100 031213201 031213312 031220121 031220200 031222110 031223223 031230012 031230331 031231132 031232202 031232313 031233230 031300131 031300223 031301300 031302020 031310212 031310320 031311331 031312113 031313102 031320301 031321011 031321233 031323022 031333010 031333303 032000013 032001203 032002232 032011011 032012133 032013202 032020221 032021112 032022020 032022313 032023003 032023331 032030102 032031033 032031310 032032322 032100331 032101000 032102012 032102223 032103120 032110021 032110233 032112111 032112200 032121213 032122301 032123030 032123322 032130032 032132130 032133113 032200002 032200220 032201122 032203333 032210103 032211020 032212001 032212310 032220312 032221300 032222102 032222230 032223111 032230110 032231323 032232211 032233302 032300303 032301113 032301332 032302201 032311221 032312120 032313013 032313300 032320123 032321101 032330321 032333233 033003102 033010321 033011000 033013223 033013330 033020110 033021301 033023032 033023121 033023200 033030220 033032021 033100323 033102211 033103310 033110130 033111012 033112313 033113001 033113122 033121023 033122332 033130231 033130302 033131100 033131333 033132003 033132222 033200033 033200300 033201001 033203213 033210022 033211113 033212132 033220011 033221202 033222320 033223231 033231312 033233020 033233103 033300120 033302133 033303031 033310111 033310200 033310332 033312301 033320102 033321210 033321322 033322000 033330013 033331002 033331223 033332112 033333311 100001033 100001310 100003302 100010032 100010210 100021201 100023012 100023233 100023321 100030231 100030303 100031112 100032013 100032100 100032222 100100020 100100213 100102011 100110100 100110322 100111211 100112132 100113002 100121023 100123113 100132210 100132323 100133031 100133120 100201101 100211113 100212030 100212202 100212321 100213122 100213333 100220121 100221000 100222313 100223220 100230002 100231320 100233312 100300201 100300323 100301232 100302112 100302220 100311131 100312003 100313300 100320102 100320230 100321011 100322123 100330110 100331213 100333022 100333103 101001000 101002303 101003220 101010123 101010301 101012200 101012332 101013131 101020212 101022221 101023100 101030010 101033021 101100122 101100331 101101103 101110233 101111330 101112022 101121121 101121313 101122102 101122230 101123001 101130130 101131200 101200032 101201120 101201333 101203023 101210013 101211322 101213000 101213311 101220103 101220330 101222131 101223202 101230111 101231212 101233233 101300300 101302202 101303011 101310132 101311100 101311221 101312320 101320223 101321020 101323303 101330001 101331332 101332030 101332113 102000001 102001332 102002211 102003010 102011311 102012002 102013222 102020033 102022103 102022330 102030312 102031233 102033003 102033320 102100200 102102301 102103121 102110011 102111101 102113203 102113310 102121032 102122222 102123020 102130023 102131010 102132231 102133302 102200131 102202020 102203102 102203313 102211220 102212111 102220010 102221323 102222302 102223331 102231132 102232223 102232310 102233201 102301013 102302333 102303130 102310303 102311202 102312213 102313001 102313123 102320120 102322021 102323111 102323232 102330222 102331031 102332200 103000203 103002122 103011333 103013211 103020132 103021210 103022311 103023030 103031020 103031101 103033133 103100033 103101110 103102131 103103212 103103320 103110220 103111021 103112103 103120310 103130232 103131002 103133223 103200311 103201030 103201221 103202113 103202332 103210133 103211300 103212023 103220001 103220222 103221312 103222233 103230323 103232011 103233100 103300022 103300210 103301123 103301302 103303003 103312032 103313313 103320333 103321321 103322110 103323122 103330121 103332212 103333330 110000132 110001120 110001301 110002030 110010320 110011231 110013102 110020100 110022211 110022333 110023131 110023222 110031010 110032312 110100031 110101222 110102102 110103332 110110232 110111013 110113110 110113201 110121230 110123033 110130211 110132003 110200003 110201213 110202201 110203310 110210021 110212210 110213032 110220030 110221012 110221331 110222300 110230112 110231103 110232121 110233000 110233223 110302022 110302133 110303321 110310002 110310313 110311212 110311330 110312301 110322013 110322232 110323210 110330101 110331200 110331323 110333302 111000033 111000202 111001313 111003322 111010112 111011210 111012233 111013221 111020303 111021130 111022022 111030220 111031032 111032011 111032123 111032300 111033110 111033331 111100101 111102223 111102310 111103002 111111202 111111311 111112030 111113123 111120020 111120332 111122201 111123300 111130013 111131333 111132131 111133203 111133312 111200123 111201111 111201300 111210131 111211023 111212001 111212113 111212222 111213302 111221220 111223132 111230232 111231122 111232020 111300110 111301021 111301132 111301203 111302000 111313111 111320321 111322112 111323233 111330022 111331301 111332210 111333320 112000223 112000311 112010133 112010302 112012121 112013031 112020210 112022202 112023013 112030000 112031113 112031322 112032230 112033211 112100012 112101201 112103030 112110003 112110321 112111022 112111300 112112212 112112333 112113132 112120233 112122111 112122320 112123221 112130202 112131220 112132100 112133323 112200333 112201130 112202321 112210200 112211010 112213103 112220023 112222031 112222122 112222213 112223110 112230011 112230120 112231002 112232303 112300320 112302231 112303101 112303213 112312130 112313312 112320001 112320132 112321033 112321311 112323022 112323330 112330331 112331121 112333010 112333133 113001211 113002103 113002220 113003112 113003330 113010023 113011012 113012310 113013301 113020031 113021300 113023203 113030130 113030201 113031223 113033002 113100221 113101000 113102322 113103303 113111133 113112011 113112200 113113020 113120103 113121120 113122023 113123331 113131321 113132213 113133230 113201022 113211121 113213013 113213231 113221003 113222010 113223101 113223212 113223323 113230033 113230300 113232202 113233021 113233332 113301333 113302121 113303032 113310122 113310203 113313100 113320220 113320312 113321102 113322303 113330113 113332132 113332311 113333222 120000022 120000233 120001102 120002131 120003123 120011011 120011203 120012122 120013230 120020301 120021032 120022010 120022223 120031333 120100202 120100321 120103101 120110303 120111123 120112330 120121312 120122120 120130113 120131022 120133221 120200220 120201231 120201303 120202013 120203002 120210211 120211020 120211332 120213131 120222101 120222212 120230100 120230322 120231210 120232133 120300011 120301121 120302203 120302310 120310200 120312111 120313012 120320023 120321103 120321220 120322321 120323202 120323313 120331030 120331311 120332332 120333001 121000221 121000332 121003013 121010000 121010213 121011323 121020230 121022113 121030311 121031003 121031120 121032102 121101033 121101211 121101302 121102012 121102200 121103330 121112133 121112321 121113220 121120110 121120323 121121000 121121132 121123103 121130032 121130300 121131223 121133011 121200001 121202232 121203133 121203312 121211103 121212300 121220012 121221021 121222333 121223213 121231201 121232110 121233030 121233222 121300020 121300212 121302331 121311002 121311310 121313021 121320101 121321322 121330123 121333112 121333333 122000103 122001021 122002033 122002110 122003202 122011232 122012201 122013300 122020011 122020222 122021213 122021320 122023112 122023333 122030132 122033223 122100111 122101230 122102221 122102332 122112013 122121331 122123130 122130210 122131001 122132030 122132123 122132311 122201000 122201223 122201311 122203210 122211031 122212120 122213022 122220113 122220332 122221102 122222200 122223003 122230233 122230301 122231013 122233111 122300002 122301133 122303023 122310110 122311122 122311211 122312302 122320203 122322012 122322131 122330313 122331100 122332320 122333231 123001130 123001322 123003000 123003311 123010111 123011220 123012031 123013103 123013332 123021123 123022002 123023021 123030302 123032121 123032233 123033210 123100100 123102313 123103022 123110312 123111102 123113323 123120122 123120201 123121013 123123233 123130003 123131330 123132112 123132220 123133131 123133202 123200330 123202102 123202211 123203031 123210002 123210223 123211212 123213310 123221230 123222322 123230010 123231032 123300131 123301012 123303110 123303201 123310033 123311113 123311331 123312222 123320232 123321001 123322213 123323300 123330211 123331203 123332000 123333013 123333321 130001223 130002021 130003212 130003333 130011321 130012303 130013001 130021020 130021313 130023110 130030133 130030310 130031302 130033200 130100110 130102233 130103122 130111031 130112010 130112222 130113320 130121221 130122032 130130203 130131000 130202123 130203100 130210033 130211102 130220231 130221130 130223302 130231111 130231222 130233013 130233321 130300103 130300330 130301002 130310221 130311023 130312312 130313213 130320000 130320111 130321122 130322201 130330031 130330212 130332230 130333132 131000210 131001122 131001331 131003103 131010022 131011033 131011111 131011300 131012212 131013120 131020001 131020320 131022203 131023023 131023312 131030121 131032130 131033232 131100023 131101001 131101232 131102111 131110200 131111213 131113112 131113301 131122021 131123211 131132122 131132303 131133133 131200130 131200203 131200321 131201012 131211231 131212323 131220313 131221301 131222000 131230302 131233001 131301323 131303222 131303310 131310311 131312131 131313003 131313332 131321032 131321110 131322302 131323121 131323200 131330100 131330233 131331013 131331220 132000120 132002300 132003032 132010331 132011130 132012223 132021121 132022132 132023230 132030213 132032010 132033122 132101112 132101321 132102031 132102202 132110030 132110113 132111332 132120022 132120300 132122003 132122210 132123102 132123313 132130221 132131103 132133212 132133330 132200310 132201302 132202101 132203011 132210322 132211313 132212012 132220202 132221211 132223133 132223320 132230003 132231030 132232021 132232232 132300021 132300232 132301210 132302122 132303331 132310101 132311233 132312321 132313020 132320013 132321303 132322100 132331201 132331312 132332033 132332111 132333002 133000030 133000101 133000313 133001200 133002012 133002231 133012113 133012322 133013010 133020211 133021332 133022033 133022120 133030112 133031311 133033221 133033303 133100002 133102330 133103013 133110121 133110333 133111003 133111310 133113032 133120223 133121101 133121212 133123000 133130011 133131132 133132201 133133110 133200122 133201233 133202003 133203301 133210100 133212230 133213111 133213202 133220020 133221031 133222112 133222221 133230131 133231120 133232333 133301111 133302223 133303230 133310012 133311030 133312102 133312211 133320130 133322022 133322331 133323103 133330322 133331021 133331300 200000023 200000112 200001130 200001221 200002010 200003103 200003232 200010133 200011111 200013320 200020211 200022033 200023313 200031210 200032002 200033301 200101031 200102022 200102201 200103310 200111220 200112123 200112302 200113030 200113112 200120010 200123003 200123222 200132331 200133213 200200001 200200320 200202312 200203131 200211323 200213210 200220203 200221013 200221232 200222120 200223021 200230132 200231100 200233330 200300100 200302321 200303013 200312012 200312231 200321030 200321101 200321312 200330020 200330233 200332122 200332310 200333202 200333323 201001302 201002120 201003033 201003111 201010030 201011201 201012103 201012321 201020003 201020332 201021012 201021131 201022202 201023230 201030022 201031330 201032213 201100021 201100110 201100203 201102133 201103000 201103332 201110312 201111102 201113323 201121211 201122320 201131003 201131322 201133131 201200230 201201011 201201132 201210223 201212122 201212313 201220031 201221321 201223101 201223312 201230310 201232012 201232231 201233220 201301222 201310331 201311013 201313020 201313233 201320130 201320201 201322022 201323113 201330212 201332302 202001020 202001313 202010322 202012023 202013102 202020101 202023022 202023133 202030300 202031001 202031122 202033110 202033221 202100323 202101212 202101300 202103113 202112010 202112131 202113021 202120220 202122100 202122233 202123301 202131223 202132032 202133333 202201121 202202232 202203012 202211022 202211233 202212300 202213120 202220122 202221130 202222003 202222210 202230030 202231302 202232321 202233103 202300213 202300332 202301103 202302031 202310221 202311132 202311310 202313212 202320011 202321023 202322112 202323320 202330003 202330131 203000220 203000301 203001032 203002212 203010012 203013031 203013113 203013200 203020123 203022001 203033312 203100330 203102303 203103011 203103102 203110122 203111130 203111311 203113232 203121202 203122030 203123210 203130000 203130133 203133201 203200010 203200202 203202033 203202100 203203323 203210003 203210321 203211112 203212211 203221303 203222132 203230101 203231222 203233002 203302111 203303231 203303300 203310110 203311121 203311203 203312000 203312333 203320322 203321133 203322013 203322221 203323311 203331012 203331320 203332103 203332230 203333021 210000333 210002123 210003211 210003300 210010121 210010230 210011103 210012032 210013010 210021330 210022200 210023023 210030011 210031022 210032221 210033112 210033203 210100113 210101320 210111002 210113133 210113321 210120001 210120122 210130323 210131130 210132300 210201000 210210310 210211120 210212011 210212303 210220133 210220212 210221221 210223322 210230220 210230302 210232233 210233311 210300021 210300312 210301131 210301223 210312113 210313101 210313220 210313332 210320110 210323231 210332001 211000001 211001112 211001321 211010311 211011000 211011223 211012330 211020010 211022323 211030231 211031202 211033313 211100032 211100211 211101200 211102122 211103013 211113222 211120223 211122033 211131012 211132311 211200322 211201231 211202002 211202333 211203301 211210033 211212200 211213121 211220111 211221102 211221213 211222021 211223000 211231110 211233023 211233212 211300303 211303133 211310213 211311022 211311333 211312031 211313300 211321230 211323012 211330000 211331123 211332013 211332232 211333221 212000232 212001002 212002011 212003131 212011110 212012222 212013123 212020120 212021212 212022030 212023001 212030103 212030321 212032132 212033033 212033302 212101333 212103322 212111011 212112102 212112230 212120302 212121123 212123232 212123310 212130111 212130330 212132201 212133020 212200221 212200300 212203230 212210020 212210101 212211003 212213333 212220313 212221200 212221332 212222301 212223223 212231131 212231320 212232010 212300010 212301301 212302023 212302100 212310112 212310323 212311030 212312203 212312311 212320222 212322121 212330032 212331102 212331213 213000022 213002000 213003223 213010100 213021220 213022111 213023102 213031030 213031121 213032012 213032303 213101101 213102210 213110010 213111221 213112003 213112120 213112332 213113213 213122321 213123022 213123130 213130102 213131113 213133031 213200031 213200112 213201133 213201311 213210123 213211032 213212131 213213001 213220002 213222103 213222312 213230211 213232330 213233200 213300230 213301003 213301110 213302313 213303122 213311312 213313033 213320300 213321011 213321323 213322233 213330120 213331231 213332322 213333310 220002100 220003001 220003322 220010013 220011310 220013333 220020021 220020130 220021112 220022231 220030102 220030223 220030331 220031000 220032113 220033032 220101010 220102212 220102333 220103231 220110201 220111233 220112311 220113022 220121300 220122103 220123132 220131121 220131313 220133320 220200012 220201033 220202121 220202230 220203313 220210031 220212000 220213103 220220323 220221002 220221131 220222332 220223110 220223201 220230111 220231301 220232202 220233122 220300003 220303112 220303330 220310120 220311102 220312033 220312322 220313211 220320302 220321333 220322011 220323000 220323121 220331023 220332220 220332303 220333133 221001031 221001210 221002203 221003132 221010303 221011122 221012002 221012111 221012220 221021020 221022300 221030200 221032021 221100123 221100301 221101130 221110320 221111113 221112100 221112231 221113001 221120011 221120232 221123313 221131331 221132210 221132323 221133033 221133102 221201022 221202311 221203100 221210010 221211302 221213112 221222130 221222222 221223032 221223123 221230002 221230120 221230213 221231133 221232101 221233303 221300233 221300310 221303323 221311131 221312301 221313202 221321100 221321212 221322003 221323220 221323331 221330132 221331010 221333022 222000000 222001111 222001233 222002022 222002301 222003120 222010211 222011012 222011331 222013003 222020032 222021302 222022013 222023200 222023321 222031130 222031323 222032212 222033101 222100013 222101032 222102310 222110002 222111303 222113110 222120112 222120333 222121101 222121222 222122031 222130021 222130100 222131211 222133230 222200110 222200331 222201201 222202133 222203021 222210203 222210312 222211100 222213301 222220001 222221010 222222111 222223212 222223330 222230222 222231112 222232313 222233000 222300101 222300220 222301312 222302211 222303203 222310232 222310300 222311001 222311223 222312020 222313313 222321113 222321231 222322323 222323033 222323102 222332002 222332110 222332331 222333011 223000011 223001023 223001300 223010132 223011101 223012010 223012323 223013222 223020320 223021203 223021311 223022122 223030033 223030110 223032332 223033020 223102001 223102223 223103133 223110103 223111000 223112202 223113121 223120131 223121021 223121110 223122211 223130012 223131232 223132130 223133311 223200303 223201120 223201213 223201332 223203111 223211011 223212233 223220100 223220221 223222020 223223013 223223302 223230230 223231003 223231310 223232031 223233223 223300113 223301030 223301221 223302200 223303002 223310022 223310311 223311210 223312112 223313320 223320010 223322032 223322101 223330001 223330323 223331111 223331302 223333100 230000031 230001312 230002222 230012020 230012112 230013131 230020303 230021233 230023002 230023221 230030322 230031013 230032330 230033123 230100221 230100300 230101123 230110032 230112001 230112130 230120213 230122322 230123111 230123333 230131102 230132033 230133010 230201110 230201321 230202203 230203332 230210202 230211012 230213023 230220311 230221103 230222010 230223230 230230000 230231231 230232022 230233101 230301011 230301230 230302102 230310010 230311301 230313122 230322223 230322300 230330121 230330313 230331332 230332211 230333003 231000020 231002013 231010110 231010233 231013021 231021113 231021200 231021322 231030301 231031221 231032032 231033000 231100333 231102030 231103223 231103311 231111010 231111121 231113103 231113330 231120000 231122101 231122212 231123031 231123122 231130112 231130220 231131233 231200101 231200212 231203113 231210300 231211001 231211222 231212133 231220023 231221310 231222303 231231020 231232223 231233322 231301213 231302110 231302221 231303302 231310002 231312023 231313210 231320103 231322311 231323232 231330011 231331101 231332200 231332333 231333130 232000133 232000302 232001100 232002210 232003201 232010200 232012101 232022123 232023010 232030012 232032003 232032231 232100001 232100122 232100230 232101131 232102103 232110310 232111033 232111201 232112213 232113000 232121320 232122221 232131022 232131110 232132011 232132302 232133121 232133203 232201013 232201330 232202000 232202112 232202323 232210011 232211123 232211210 232212331 232213132 232213221 232220131 232221021 232222032 232223100 232230201 232230332 232231311 232232120 232233031 232300311 232301202 232302233 232303030 232303123 232310130 232311111 232311322 232320020 232321012 232322001 232323211 232330210 232331000 232331133 232332222 232333112 232333301 233001010 233002121 233002333 233003003 233003130 233010001 233011133 233011211 233011320 233012203 233020013 233020222 233022310 233032100 233033111 233033233 233100111 233101313 233102232 233103020 233110023 233110212 233111302 233120120 233120301 233121032 233122113 233122200 233123312 233131001 233133300 233200223 233201102 233202011 233210231 233212002 233212110 233213030 233213313 233220210 233220333 233221000 233221111 233223203 233223321 233230021 233230113 233232212 233232301 233300000 233300132 233301022 233301331 233302320 233303101 233303212 233310303 233311100 233311232 233313011 233320031 233321201 233323023 233323110 233330202 233330330 233332010 233332131 233333032 233333220 300002333 300010311 300011123 300011300 300012102 300013033 300020103 300023200 300031332 300033122 300100111 300100332 300101200 300103221 300110003 300110230 300111110 300120313 300121320 300122202 300131101 300133232 300200133 300200222 300201302 300202211 300203120 300211212 300212022 300220032 300221111 300222330 300223102 300223323 300230321 300232103 300301122 300302030 300303311 300310320 300311233 300312210 300313113 300313222 300320001 300322132 300322303 300330013 300330302 300331000 300332111 300333130 301000113 301002022 301002230 301010231 301011002 301020021 301021232 301023211 301031033 301031110 301100012 301101030 301101321 301102101 301103123 301111222 301112211 301112300 301113032 301120133 301121100 301122003 301122312 301123330 301130221 301132132 301133311 301202010 301202223 301202331 301203002 301210020 301210332 301211130 301211203 301223110 301231001 301232320 301300031 301301201 301303100 301303212 301310111 301321213 301321331 301322200 301323322 301330330 301331022 301331303 301332233 302000330 302001101 302002302 302003031 302003112 302003223 302010010 302013301 302020002 302021013 302022120 302032021 302032133 302033230 302101011 302111133 302112332 302113111 302120030 302120121 302123103 302123212 302130331 302132013 302133000 302200203 302202122 302210001 302210112 302211321 302212033 302213202 302220320 302221031 302222201 302223233 302230100 302230232 302231211 302233322 302300123 302301032 302302110 302303020 302310022 302310200 302311003 302311120 302321300 302322311 302323221 302331113 302333012 302333203 303000100 303003013 303003321 303010222 303012233 303021000 303021221 303022131 303023310 303030032 303031213 303032201 303032323 303100001 303102032 303103333 303110113 303111323 303112020 303113100 303121012 303121230 303122122 303122213 303123023 303130303 303131031 303132110 303133121 303200121 303201003 303201232 303203200 303203312 303210210 303211101 303213011 303213132 303220013 303220302 303221123 303231333 303233030 303300112 303301010 303301131 303302222 303311211 303311330 303312321 303320020 303320231 303323101 303330223 303330311 303331202 303332332 303333210 310000013 310001203 310002232 310011001 310011130 310011312 310012111 310012223 310013022 310020112 310020233 310022020 310030300 310033231 310100310 310101121 310103012 310103230 310110213 310112000 310112122 310113031 310120221 310121133 310123302 310130020 310130103 310131112 310131223 310132321 310133313 310200102 310201220 310202131 310202322 310210333 310221303 310223213 310231021 310231310 310232002 310233133 310301332 310303033 310303202 310310023 310310100 310311321 310321201 310322331 310323103 310323320 310330230 310331120 310332010 310332203 311000331 311002003 311002121 311003032 311003213 311010120 311012201 311013133 311013310 311022302 311023220 311031101 311032112 311032333 311100000 311101022 311102231 311110021 311110330 311111003 311111132 311112012 311112323 311113101 311120301 311122210 311123113 311131300 311132202 311133122 311203103 311203320 311211011 311212102 311213230 311221131 311221322 311222232 311222313 311223033 311223201 311230010 311230121 311230203 311300232 311302301 311311110 311311231 311312020 311313203 311320011 311321023 311322221 311330102 311330323 311331312 311332130 311333031 312000021 312000200 312001033 312002323 312011221 312012013 312012100 312013212 312021022 312021111 312022310 312030313 312032301 312033121 312101103 312102002 312102120 312102311 312110123 312111210 312111331 312113303 312122203 312131032 312133110 312200030 312200312 312201001 312203111 312203222 312212211 312212320 312213000 312220103 312220231 312222130 312223012 312223321 312230022 312231233 312232200 312233102 312233330 312300003 312301112 312301230 312303300 312311101 312311313 312312232 312313011 312313122 312320333 312321010 312322113 312330212 312333023 313002130 313003020 313003101 313010211 313010303 313011232 313012002 313020332 313022222 313023123 313031331 313032210 313033011 313033322 313101233 313102013 313103132 313110131 313110202 313120033 313120110 313121211 313122001 313122330 313131010 313132022 313132133 313200023 313200201 313201330 313202303 313210012 313211223 313211302 313213120 313221020 313221112 313230132 313231100 313232231 313233003 313300133 313300322 313302031 313303211 313310030 313312103 313313323 313320213 313321121 313322012 313323230 313330021 313331033 313333112 313333301 320000030 320000101 320002012 320002221 320011222 320012200 320012332 320013121 320020322 320021003 320021331 320031111 320032033 320032120 320033010 320033303 320101001 320101322 320102130 320111131 320113102 320120330 320121210 320122111 320123123 320130002 320130233 320130311 320201113 320202301 320203132 320210302 320212213 320213001 320213330 320220010 320221100 320222021 320223222 320223311 320231012 320232323 320233200 320300110 320300333 320301211 320303022 320311010 320311303 320312002 320320131 320322233 320323332 320330201 320331132 320333212 321001123 321003021 321003110 321010032 321011211 321011330 321020013 321022101 321023122 321031302 321032000 321032222 321033233 321102020 321103003 321103131 321110203 321111120 321113333 321120102 321121233 321122032 321123221 321130322 321131212 321132113 321132330 321200210 321201101 321202033 321210100 321210321 321213013 321220223 321221030 321221202 321222112 321223300 321233211 321233332 321300103 321301313 321302132 321310001 321310230 321312223 321313312 321320320 321321111 321322010 321323133 321330012 321331200 321332321 321333101 322000213 322001132 322001303 322010202 322011000 322012131 322013323 322020110 322022211 322030123 322031031 322031210 322032203 322033002 322033311 322100022 322100231 322100300 322101110 322103312 322110101 322112021 322112112 322113030 322113222 322122000 322122322 322123011 322131102 322131333 322133201 322200011 322200323 322201212 322202100 322210133 322210220 322211023 322211111 322212010 322212303 322213231 322221221 322221313 322222123 322222331 322223020 322223101 322230310 322231120 322232001 322232132 322301320 322302121 322302202 322303001 322303113 322310013 322311332 322313100 322321002 322321130 322323210 322323303 322330000 322330111 322331222 322331301 322332230 322332312 323000122 323000310 323001201 323002111 323003302 323010020 323011033 323011110 323011321 323013012 323020001 323020133 323022030 323022312 323023100 323023213 323030231 323033330 323101311 323103010 323110011 323111022 323111213 323112310 323113301 323121103 323123112 323123320 323132101 323133032 323200000 323200233 323201021 323210313 323211200 323212032 323212121 323213322 323220022 323220111 323221301 323222003 323222210 323223333 323230103 323231131 323232300 323233110 323300032 323300301 323301100 323302023 323302330 323303220 323310102 323310221 323313003 323313111 323313232 323320200 323321223 323321310 323322120 323323031 323330130 323331020 323332011 323333123 330000002 330001231 330002110 330003011 330003320 330010021 330011032 330011210 330020220 330022013 330022122 330023232 330023301 330031100 330031323 330032202 330100033 330101212 330103000 330103113 330103331 330110120 330110301 330112023 330113312 330121311 330123130 330130131 330133211 330200020 330202313 330203210 330211201 330211322 330212232 330220101 330221023 330222133 330230011 330230303 330232220 330232331 330233112 330300213 330301101 330302003 330310133 330312011 330313200 330320022 330321031 330321113 330321302 330322212 330323010 330332032 330332123 330333021 330333333 331000322 331001310 331002133 331003200 331010101 331012320 331013222 331013303 331020030 331020202 331022231 331030211 331032023 331033012 331100201 331101102 331102222 331112110 331112233 331113011 331120111 331121013 331121220 331121332 331122123 331123002 331130003 331131021 331131130 331133320 331200013 331201031 331202300 331203122 331203233 331211112 331212003 331212221 331220132 331222022 331223212 331223331 331230222 331231213 331232030 331232111 331233100 331301120 331302012 331303111 331310122 331311103 331312202 331313030 331313321 331320210 331321000 331322033 331323313 331330020 331330113 331331232 331332001 331332310 331333223 332000111 332001012 332002030 332003313 332012022 332013110 332013332 332020323 332021103 332022200 332023021 332030001 332030130 332031020 332033300 332100010 332101023 332102333 332103101 332103232 332110132 332110211 332111100 332121001 332121122 332122131 332123223 332130200 332130312 332131231 332133033 332200301 332201133 332201200 332202231 332203003 332210121 332210330 332211002 332212113 332213311 332220000 332220213 332221110 332221232 332222011 332231101 332233010 332233123 332300100 332303322 332310031 332310223 332310302 332311212 332312000 332320112 332320201 332321321 332322332 332323120 332331011 332331330 332332102 332332213 332333131 333000212 333001113 333002001 333010123 333011102 333013201 333020300 333021011 333021130 333022321 333030010 333030333 333031003 333031222 333033120 333100103 333100220 333101300 333102112 333110000 333110322 333111111 333112031 333113133 333113210 333120021 333120232 333122010 333123303 333130213 333131123 333132230 333132311 333133102 333200110 333200332 333201211 333201323 333202202 333203131 333210203 333211010 333211331 333212312 333213300 333222100 333223001 333223113 333223220 333230002 333230320 333232013 333232122 333233232 333300011 333301203 333301312 333302210 333310310 333311001 333311220 333312130 333313022 333320003 333321333 333322111 333323202 333330101 333331110 333332221 333332303 333333000\par}

\paragraph{$E_4(6,4) \ge 30$, 30 words.} Stabiliser in $S_4\times C_2$: $\{\mathrm{id},\ (0\,1)(2\,3),\ ((2\,3),\mathrm{rev}),\ ((0\,1),\mathrm{rev})\}$ (diagonal, order 4). sha256 {\footnotesize\texttt{fc6fb24268384a77f148646c7760bb3049b57258d28386310f392f89675a270f}}.

{\small\ttfamily\sloppy 000333 001011 003122 013210 020202 021130 032231 033003 102301 110100 111222 112033 122112 123320 130021 131313 200220 201103 212121 220311 222000 223233 230132 303030 310012 311331 321023 331200 332322 333111\par}

\paragraph{$E_4(7,4) \ge 72$, 72 words.} Stabiliser in $S_4\times C_2$: the order-8 diagonal group generated by $((0\,1),\mathrm{rev})$, $((2\,3),\mathrm{rev})$ and $(0\,2)(1\,3)$. sha256 {\footnotesize\texttt{85cb1ceb68de50ccc1566cddddc8240d28b328cbb2b688e8cc0520bc74aeb391}}.

{\small\ttfamily\sloppy 0000022 0001210 0012221 0020300 0022132 0033333 0110212 0120033 0131000 0203223 0210130 0221103 0222011 0233112 0303302 0311313 0323231 0330220 1001303 1020111 1031122 1103330 1110301 1111133 1122222 1131211 1133023 1200202 1212213 1221331 1232320 1301021 1312332 1322003 1330012 1333100 2000233 2003321 2011330 2021001 2032312 2101013 2112002 2121120 2133131 2200310 2202122 2211111 2222200 2223032 2230003 2302211 2313222 2332030 3003113 3010102 3022020 3030031 3100221 3111322 3112230 3123203 3130110 3202333 3213300 3223121 3300000 3311201 3313033 3321112 3332123 3333311\par}

\paragraph{$E_4(8,4) \ge 200$, 200 words.} Stabiliser in $S_4\times C_2$: the order-8 diagonal group generated by $((0\,1),\mathrm{rev})$, $((2\,3),\mathrm{rev})$ and $(0\,2)(1\,3)$. sha256 {\footnotesize\texttt{2083125e16976bdc2c73a6de79fef82e798668ebb92c84ac32b9c3a4c148d251}}.

{\small\ttfamily\sloppy 00003213 00012232 00022100 00032021 00103332 00111013 00130122 00133000 00200300 00210211 00220023 00231233 00311200 00321032 00333311 01002333 01023010 01031212 01113233 01120202 01132110 01211310 01221122 01230032 01233201 01300210 01310301 01320133 01331023 02000120 02011030 02021323 02101021 02123111 02130313 02212012 02223330 02232220 02302003 03001131 03020312 03030223 03100030 03102222 03111120 03222231 03233322 03312113 03313002 03332330 10002322 10023001 10031313 10113222 10120303 10132101 10201210 10211301 10220132 10231022 10300201 10321123 10322310 10330033 11000102 11012223 11021033 11022111 11103323 11112302 11123130 11133011 11200311 11222200 11230123 11301300 11311211 11320322 11331132 12000031 12011121 12013333 12110020 12121332 12131203 12202113 12203002 12223221 12322233 12333320 13010130 13021202 13032000 13100121 13111031 13130232 13213112 13303103 13323331 13332221 20001112 20010002 20030230 20120221 20203101 20222302 20233212 20301333 20312131 20323203 21000013 21011100 21110112 21130331 21131220 21202130 21212001 21223313 21320000 21322212 21333302 22002201 22013011 22022122 22032033 22103210 22111133 22133022 22200322 22210203 22221031 22230010 22311222 22312300 22321110 22333231 23003300 23011023 23012210 23033132 23102311 23113201 23122032 23132123 23201232 23213030 23220111 23302020 23310332 23331011 30001003 30020331 30021220 30100011 30111102 30222213 30231111 30233303 30303110 30313021 30332202 31031330 31101113 31110003 31121321 31203020 31210222 31232312 31312010 31322303 31333213 32002310 32013200 32023032 32033123 32100132 32103301 32112211 32122023 32201223 32213131 32220100 32302121 32310323 32331000 33000022 33012301 33022133 33102100 33113310 33123122 33133033 33200333 33203211 33222320 33230001 33301312 33311233 33321101 33330120\par}

\paragraph{$E_4(9,4) \ge 520$, 520 words.} Stabiliser in $S_4\times C_2$: the order-8 diagonal group generated by $((0\,1),\mathrm{rev})$, $((2\,3),\mathrm{rev})$ and $(0\,2)(1\,3)$. sha256 {\footnotesize\texttt{f48053ee681fc5791c5a4f367cecd8c5bbd3d32c45dd1ffe10c194e5f5746569}}.

{\small\ttfamily\sloppy 000000211 000011031 000023301 000030333 000101323 000112300 000121012 000202222 000203000 000222130 000331122 000333230 001100102 001133203 001220020 001232011 001310231 001322223 002003110 002013022 002111111 002112232 002203321 002211333 002232123 002300013 002311000 002320202 003001332 003021221 003120110 003123033 003213112 003300320 003333311 010003121 010022033 010103032 010112221 010210230 010300001 010302113 010323222 011010110 011031200 011110322 011113101 011200300 011201221 011222112 011233133 011302202 012011132 012033313 012100212 012121031 012223023 012232032 012322120 012333201 013010023 013101122 013113330 013132323 013203210 013220103 013321111 013331302 020001002 020102312 020121220 020230112 020311101 020332023 021003322 021032303 021103113 021111202 021130021 021312130 021320012 021333000 022000223 022012001 022021311 022120133 022133120 022200101 022211210 022222221 022310303 022331113 023003231 023013010 023202333 023211032 023222000 023301203 023312211 023330131 030002132 030031223 030033100 030133321 030222203 030301033 030323131 031001111 031020201 031112331 031211123 031221330 031231002 031313013 031332310 032002200 032022113 032030302 032103011 032220122 032311321 032332233 033023312 033112102 033120232 033131030 033200031 033210213 033233303 033300112 100001233 100002010 100101001 100120311 100213313 100310330 100311211 100322022 100333003 101003330 101012123 101112030 101133122 101211110 101213002 101232333 101301321 102002221 102010033 102023232 102101132 102220213 102230000 102312301 102331012 103011303 103030120 103100023 103122202 103222310 103233031 103323123 103332132 110011013 110022312 110201320 110233332 110323100 110331131 111003211 111010232 111030103 111100120 111111300 111121222 111132210 111220033 111222321 111312111 111313333 111333021 112031001 112032122 112110223 112130330 112211231 112222200 112302003 113000000 113003323 113102133 113112001 113200111 113211102 113231313 113300222 113312230 113323032 120003220 120110000 120131310 120202102 120223201 120300032 120320113 120330221 121022230 121113023 121120332 121122011 121210122 121232020 121333312 122003013 122020121 122031323 122132203 122211003 122301302 122311120 122322212 123012100 123113311 123121213 123133002 123200230 123223322 123331033 130000313 130012002 130021130 130112233 130123212 130203021 130222111 130231103 131013203 131030331 131110113 131200010 131223132 131321003 132102101 132112320 132203300 132210312 132221020 132300123 132313222 132333111 133022031 133031022 133103110 133111332 133130200 133201212 133220002 133300301 133311010 133333330 200000003 200022323 200033032 200113331 200132121 200203133 200222001 200230223 200302311 200311302 201000222 201020111 201033210 201112313 201123021 201130033 201221013 201231232 202012330 202110201 202133323 202223220 202303002 202320130 203102230 203111222 203130312 203210121 203221100 203312203 203321331 203333020 210002300 210110011 210133103 210200331 210212120 210220022 210321233 211011121 211022213 211032031 211122330 211201130 211302010 211313212 211330320 212000021 212101313 212123211 212211322 212213001 212220310 212311103 213003112 213013220 213033301 213110132 213131231 213202023 213223333 213330113 220010301 220021103 220033111 220102020 220122231 220133222 220221332 220231200 220330010 220333333 221031330 221111133 221122102 221203003 221223110 221301211 221302332 222000312 222020000 222021222 222111012 222113300 222201123 222212111 222222033 222233213 222303230 222323101 222330122 223002202 223010233 223100001 223132013 223311021 223322320 230001201 230010210 230100302 230111023 230211131 230233310 230303213 230322030 231002321 231021032 231103333 231113120 231232201 231310101 231323300 231331112 232032012 232101000 232120331 232122223 232200211 232221303 232321210 232330003 233000330 233011311 233022122 233023003 233120020 233213022 233232332 233331323 233332100 300033221 300100030 300123120 300133302 300202303 300213101 300221231 300310021 301001100 301022012 301113211 301230322 301303031 301311220 301331133 302001023 302020320 302102331 302112003 302122210 302221002 302313132 302332222 303010202 303032300 303111130 303200012 303300233 303302110 303331201 310003202 310021122 310032130 310111333 310122301 310131000 310320323 310330102 311002220 311023030 311111112 311122123 311133232 311200213 311213200 311312022 311321332 311333110 312000333 312013321 312021203 312203312 312222131 312230220 312301030 312330011 313001310 313022232 313103221 313212113 313231021 313332331 320002031 320012222 320113230 320130123 320201010 320220003 320232211 320323310 321000132 321011213 321101301 321110310 321212302 321233121 321300020 321322201 322031131 322100200 322111221 322132112 322133033 322220232 322223011 322302133 322323223 323010111 323031220 323033332 323123103 323221112 323230301 323311300 323330212 330000022 330033013 330120221 330210300 330213223 330312112 330332001 331013131 331022333 331033320 331101210 331122000 331130012 331221101 331222222 331320311 331330223 332011110 332023102 332101322 332113313 332200130 332233231 333000103 333002211 333111203 333130333 333131111 333212321 333221033 333232010 333303000 333310032 333322302 333333122\par}

\paragraph{$E_4(7,5) \ge 20$, 20 words.} Stabiliser in $S_4\times C_2$: Klein four-group $V_4$ on symbols, no reversal. sha256 {\footnotesize\texttt{74f19f275251e84fc6cde1a1dc58efedefb314f53d0c830e5021537f7438a553}}.

{\small\ttfamily\sloppy 0003332 0031211 0110012 0222202 0311320 1001103 1112223 1120300 1200231 1333313 2000020 2133102 2213033 2221110 2332230 3022013 3111131 3223321 3302122 3330001\par}

\paragraph{$E_4(8,5) \ge 44$, 44 words.} Stabiliser in $S_4\times C_2$: Klein four-group $V_4$ on symbols, no reversal. sha256 {\footnotesize\texttt{b777627b5733905edf22910b860ee33ed56227f5f2e18ba73d090b27b20fdc99}}.

{\small\ttfamily\sloppy 00033021 00202003 00221212 01000120 01231133 01312201 02111110 02132322 02223331 03012032 03323000 10111031 10203310 10320022 11122130 11313112 11330303 12103123 12232111 13000001 13023233 13332220 20001113 20310100 20333332 21101222 21230210 22003030 22020221 22211203 23013311 23130023 23222302 30010333 30321301 31110002 31201011 31222223 32021132 32102200 32333213 33112121 33131330 33300312\par}

\paragraph{$E_4(9,5) \ge 100$, 100 words.} Stabiliser in $S_4\times C_2$: the order-8 diagonal group generated by $((0\,1),\mathrm{rev})$, $((2\,3),\mathrm{rev})$ and $(0\,2)(1\,3)$. sha256 {\footnotesize\texttt{f0647e8c174766334d8c526b705cc53cc77a5d21fb6a959e1f7ec85dd51eea4a}}.

{\small\ttfamily\sloppy 000000131 000013202 000302000 000321221 002211031 002322332 003310323 010230022 011001033 011331302 012232201 013010210 013120301 020022230 022023111 022203330 023112013 023330030 030111111 030123333 031002211 031211223 032133012 033200113 033221002 100110122 100220213 101321133 102031210 102101301 103323310 111102313 111111020 111213111 111230330 112201232 113233223 113300120 120113300 120300332 121000000 121032222 122311002 122330113 123022103 131133321 132003102 132221121 133132000 133312221 200021112 200201333 201112212 201330231 202200012 210311230 211003220 211022331 212301111 212333333 213033001 213220033 220033213 220100110 221132101 222103003 222120222 222222313 222231020 230010023 231232032 231302123 232012200 233113120 233223211 300112331 300133220 301200321 302122110 302331122 303210000 303222222 310003303 310221320 311130003 311310222 313311103 320213032 320323123 321101132 322002031 322332300 323103311 330023010 331011001 331122302 333012112 333031333 333320131 333333202\par}

\paragraph{$E_4(8,6) \ge 16$, 16 words.} Stabiliser in $S_4\times C_2$: $V_4 \times \langle\mathrm{rev}\rangle$ (order 8). sha256 {\footnotesize\texttt{773280cf9dc013ce9337801d25485c142dfa0ca8d2f17d10faa71c9f575d903f}}.

{\small\ttfamily\sloppy 00033000 01131332 02100120 03332221 10020223 11122111 12223330 13011031 20322302 21110003 22211222 23313110 30001112 31233213 32202001 33300333\par}

\paragraph{$E_4(9,6) \ge 32$, 32 words.} Stabiliser in $S_4\times C_2$: the order-8 diagonal group generated by $((0\,1),\mathrm{rev})$, $((2\,3),\mathrm{rev})$ and $(0\,2)(1\,3)$. sha256 {\footnotesize\texttt{f034a20a9dac757148b2ea2653eca03b60a5f233aa60ddde872f667e20c8ef99}}.

{\small\ttfamily\sloppy 000000322 000111000 001122112 001333333 021031021 022002233 030231113 033301202 110033003 110222222 111000111 111111233 121320002 122210313 130120130 133113322 200220011 203213203 211123020 212013331 222222100 222333222 223111111 223300330 300032131 303102220 311331100 312302312 332000000 332211221 333222333 333333011\par}

\end{document}